\input{aipcheck}

\documentclass[
    ,final            
  ]
  {aipproc}

\layoutstyle{8x11single}

\newcommand{\Saclay}{DAPNIA, Centre d'\'Etudes Nucl\'eaires de Saclay (CEA-Saclay), Gif-sur-Yvette, France}
\newcommand{\APC}{APC, Université Paris 7 Denis Diderot, Paris, France}
\newcommand{\LAL}{Laboratoire de l'Acc\'el\'erateur Lin\'eaire, Orsay, France}

\newcommand{\Ioannina}{University of Ioannina, Ioannina, Greece}
\newcommand{\Dortmunt}{University of Dortmunt, Dortmunt, Germany}
\newcommand{\Thessaloniki}{Aristotle University of Thessaloniki, Greece}
\newcommand{\Athens}{National Center for Scientific Research ``Demokritos'', Athens, Greece}

\begin{document}

\title{NOSTOS: a spherical TPC to detect low energy neutrinos}



\author{S. Aune}{
  address={\Saclay}
}
\author{P. Colas}{
  address={\Saclay}
}
\author{J. Dolbeau}{
  address={\APC}
}
\author{G. Fanourakis}{
  address={\Athens}
}
\author{E. Ferrer Ribas}{
  address={\Saclay}
}
\author{T. Geralis}{
  address={\Athens}
}
\author{Y. Giomataris}{
  address={\Saclay}
}
\author{P. Gorodetzky}{
  address={\APC}
}
\author{G. J. Gounaris}{
  address={\Thessaloniki}
}
\author{I. G. Irastorza  \footnote{attending speaker: Igor.Irastorza@cern.ch}
}{
  address={\Saclay}
}

\author{K. Kousouris}{
  address={\Athens}
}
\author{V. Lepeltier}{
  address={\LAL}
}
\author{T. Patzak}{
  address={\APC}
}
\author{E. A. Paschos}{
  address={\Dortmunt}
}
\author{P. Salin}{
  address={\APC}
}
\author{I. Savvidis}{
  address={\Thessaloniki}
}
\author{J. D. Vergados}{
  address={\Ioannina}
}

\begin{abstract}
A novel low-energy ($\sim$few keV) neutrino-oscillation experiment
NOSTOS, combining a strong tritium source and a high pressure
spherical Time Projection Chamber (TPC) detector 10 m in radius
has been recently proposed. The oscillation of neutrinos of such
energies occurs within the size of the detector itself,
potentially allowing for a very precise (and rather
systematics-free) measure of the oscillation parameters, in
particular, of the smaller mixing angle $\theta_{13}$, which value
could be determined for the first time. This detector could also
be sensitive to the neutrino magnetic moment and be capable of
accurately measure the Weinberg angle at low energy. The same
apparatus, filled with high pressure Xenon, exhibits a high
sensitivity as a Super Nova neutrino detector with extra galactic
sensitivity. The outstanding benefits of the new concept of the
spherical TPC will be presented, as well as the issues to be
demonstrated in the near future by an ongoing R\&D. The very first
results of small prototype in operation in Saclay are shown.
\end{abstract}

\maketitle


\section{Introduction}

Nowadays there is a compelling evidence that neutrinos change
flavor as they propagate. Appearance or disappearance of neutrinos
has been solidly proved in experiments looking at neutrinos of
either extraterrestrial~\cite{Fukuda:1998mi} (solar and
atmospheric) or terrestrial~\cite{Ahn:2002up} (reactor or
accelerator) origin.
The neutrino mixing, which provokes the oscillation of the flavor
change probability along the propagation of the neutrino, is
invoked to explain the observed appearance or disappearance.


The atmospheric neutrino oscillation data
\cite{Fukuda:1998mi,Fukuda:2000np} strongly suggest that $\nu_\mu$
oscillate into $\nu_\tau$ with maximal mixing angle ($\theta_{\rm
atm} \sim \pi/2$) and a corresponding mass squared difference of
$\Delta m^2_{23}\simeq 3\times10^{-3} eV^2$. Results from
accelerator neutrinos support this interpretation
\cite{Ahn:2002up}. On the other hand, the solar neutrino data
\cite{Ahmad:2002jz,Fukuda:2001nj,Fukuda:2001nk} could be explained
by an oscillation of $\nu_e$ into $\nu_\mu$ and/or $\nu_\tau$ with
a non maximal --but large-- mixing angle ($\theta_{\odot} \sim
\pi/3$) and a mass squared difference of $\delta m^2_{21}\simeq
7\times10^{-5} eV^2$, which has been recently supported by
evidence of disappearance of reactor antineutrinos
\cite{Eguchi:2002dm}. The third mixing angle completing the
standard three neutrino oscillation scheme is not known but it is
constrained to be quite small $\theta_{13} < \pi/6$ \cite{pdb}.
The determination of this parameter is the remaining question to
complete our understanding of the leptonic mixing and, moreover,
it will open the way to study the CP-violating effects in the
neutrino sector \cite{pdb}.

Such small mixing angle could have measurable consequences in
experiments involving electron neutrinos but sensitive to the
oscillation length driven by the large squared-mass gap $\Delta
m^2_{23}$, i.e., the smaller oscillation length. In fact, for a
detector close enough to the neutrino source the contribution from
the larger oscillation length to the disappearance oscillation
probability is negligible, and therefore it is driven only by
$\theta_{13}$ and $\Delta m^2_{23}$ in the following way:

\begin{equation}\label{oscillation}
P(\nu_e \rightarrow \nu_e) = 1-\sin^2 2 \theta_{13}
\sin^2\pi\frac{L}{L_{23}}
\end{equation}

\noindent where $L$ is the distance between source and observed
interaction and $L_{23}$ is the oscillation length related to the
neutrino energy $E$ in the way: $L_{23}=2\pi E_\nu / (\Delta
m^2_{23})$. While for reactor and accelerator neutrinos, this
length is of the order of $\sim1$ km or $\sim100$ km respectively,
we want to point out the fact that in the case of very low energy
(few keV) neutrinos, like those emitted by a tritium source, the
oscillation length $L_{23}$ is only 13 m.

The NOSTOS concept proposes the detection of such low energy
neutrinos by using a large spherical TPC of about 10 m in radius,
surrounding the tritium source, so the whole oscillation would
occur within the detector volume. That could allow the
observation, for the first time, of the space oscillating
signature and the determination of the smaller neutrino mixing
angle $\theta_{13}$, completing our knowledge of the leptonic
mixing scheme \cite{Giomataris:2003bp}. The fact that the whole
oscillation occurs inside the detector volume allows us to highly
reduce systematic effects due to backgrounds or to bad estimates
of the neutrino flux, which is the main worry in most neutrino
experiments, where the interaction rate is measured at a single
space point.

\section{The NOSTOS detector: the spherical TPC concept}

\begin{figure}
  \includegraphics[height=.4\textheight]{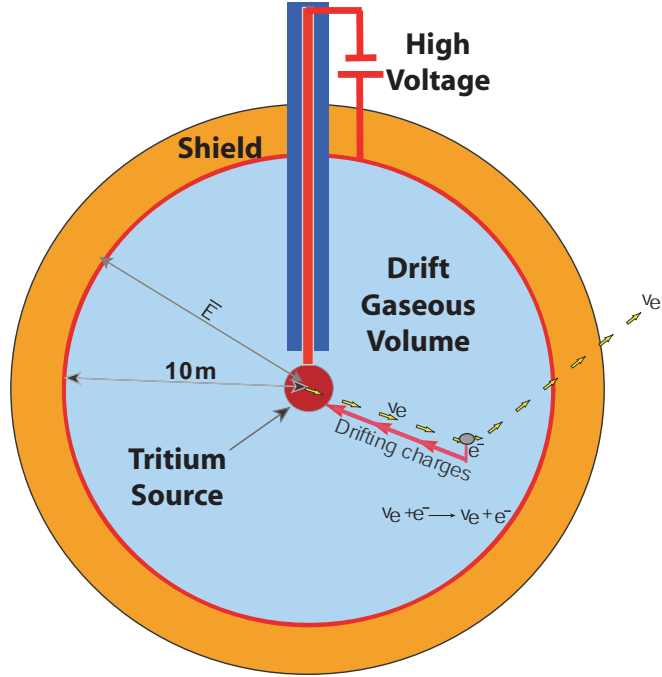}
 \caption{Illustration of the proposed NOSTOS experiment, showing
 the spherical TPC surrounded by an appropriate shield.
The neutrino source and the detector are located in the center of
the curvature of the sphere. Inside the inner sphere, the tritium
source is surrounded by the shield and cooling system and by
several Micromegas flat detectors.
 \label{NOSTOSscheme}}
\end{figure}


The NOSTOS principle of detection is based on the concept of the
spherical TPC, which has several interesting potential advantages,
as will be shown later on. It consist of 2 concentric spheres, the
external one, 10 m of radius, at ground and the inner one, 50 cm
of radius, at high potential (Fig. \ref{NOSTOSscheme}). The
tritium source is located inside the central sphere. Neutrinos
escaping radially from the center of the sphere interact in the
gas mainly by electron elastic scattering. The ionization charges
produced in the interaction drift towards the center and are
collected by an adequate gaseous readout, which covers the surface
of the inner sphere (possibly composed by a series of flat
segments).
The preferred choice for the readout devices are Micromesh Gaseous
Structures (Micromegas \cite{Giomataris:1995fq}) due to the high
precision, fast response and excellent energy resolution. High
efficiency for detecting single electrons have been proved with
Micromegas \cite{Derre:1999wh} even at high pressures
\cite{Gorodetzky}. In addition, Micromegas readout is currently
being used for solar axion detection in the CAST experiment
\cite{Andriamonje:2004hw,Andriamonje:2004hi} where a great
stability and ability to reject background events has been
achieved.

This novel approach is radically different from all other neutrino
oscillation experiments in that the neutrino source and the
detector are located in the same vessel; it is then possible to
measure the neutrino interactions continuously as a function of
the distance source-interaction point, and directly observe the
oscillation length that is fully contained in the detector. Indeed
we expect a counting rate oscillating from the centre of curvature
to the depth of the gas volume, i.e. at first a decrease, then a
minimum and finally an increase.
Fitting such an observed curve will provide all the relevant
parameters of the oscillation by a single experiment.

The radial coordinate of the interaction point is measured by
inspecting the time pattern of the charge pulse detected at the
center of the sphere which temporal extension is determined by the
longitudinal diffusion of the ionization cloud and therefore by
the distance drifted. In fact, for our spherical TPC, the electric
field at a distance $r$ from the center is given by:

\begin{equation} \label{Efield}
 E(r) = \frac{R_2 R_1 V_0}{R_2-R_1}\frac{1}{r^2}
\end{equation}

\noindent where $R_1$ and $R_2$ are the inner and outer sphere
radius and $V_0$ is the applied voltage. At low electric fields
the drift velocity $v_d$ is roughly proportional to the electric
field and the longitudinal diffusion coefficient depends on $E$ as
$D_L\sim 1/\sqrt{E}$.
Hence the spherical geometry of the electric field enhances the
longitudinal dispersion occurred after a given drifted distance
$r$, 
with respect to the one for a homogeneous field.
First calculations show that a precision better than 10 cm, which
would largely satisfy the needs of the NOSTOS concept, can easily
be achieved. Preliminary results without Micromegas regarding the
experimental demonstration of this strategy by the first prototype
actually running in Saclay are presented in section 4.

The use of Micromegas will improve those results in two ways.
First, the fast pulses then available (rise times of $\sim$1 ns)
will allow a better measurement of the temporal dispersion of the
signal; and second, information from transversal dispersion will
in also be available, by means of an appropriate pixelization of
the Micromegas readout.

\vspace{0.5cm}

In general, the use of a spherical TPC detection scheme presents
the following advantages for the proposed experiment:

\begin{itemize}
  \item The spherical geometry naturally focuses a large drift volume
  into a small amplifying
detector with only a few read-out channels. It is the most
cost-effective way of instrumenting a large detector volume with a
minimum of front-end electronics. Such approach simplifies the
construction and reduces the cost of the project.
  \item The placement of the neutrino source in the center of the
  detector provides a close to $4\pi$ acceptance. The spherical detector
  geometry optimizes the detection efficiency per unit of volume for a
  given flux of outgoing neutrinos.
  \item The spherical drift enhances the relation of longitudinal
  diffusion with drifted distance, as have been shown above,
  improving the resolution in the determination
  of the point of interaction, a key issue of the project.
  \item The ratio of external surfaces (and therefore external materials)
  over detector volume is optimized for
  a spherical geometry, therefore allowing a lower background per unit volume
  due to external surface or material contaminations. In addition,
  in a spherical geometry the thickness of material needed to hold
  the gas is minimum, further reducing possible sources of
  background.
    \item Large drift volumes can be
  built without the use of a field cage, unlike cylindrical TPCs.
  In addition, the symmetry of the design and the compact placement
  of the readout in the center of the sphere may provide a lower
  sensitivity to electronic noise (in fact, the outer sphere
  acts as a perfect Faraday cage to the inner electrode).

\end{itemize}


\section{NOSTOS experimental challenges and expected sensitivity}

The NOSTOS concept relies not only on the detection, for the first
time, of very low energy neutrinos from a tritium source, but on
the measurement of the expected oscillation in that detection rate
along the detector volume. We are confident, given the present
status of the TPC technology, and the potential of the spherical
TPC concept exposed above, on the feasibility of the project.
However, an R\&D program to assess each of the experimental key
issues on which the project relies, as well as to experimentally
demonstrate them, is already ongoing. The main experimental issues
can be summarized as follows:

\begin{itemize}
\item {\bf Energy threshold:} Neutrinos from tritium decay have a
maximum energy of 18.6 keV, therefore the maximum energy of the
recoiling electron is of only 1.27 keV. An energy threshold of the
order of $\sim100$ eV is desirable to have a good sensitivity.
Single electron efficiency has been achieved by Micromegas readout
devices \cite{Derre:1999wh}, but low threshold operation with
large volumes like that of NOSTOS is being checked experimentally,
as part of the ongoing development phase. \item {\bf Background:}
Although absence of background is not a requisite for the NOSTOS
scheme, as runs without the source can be performed to determine
the background, it is true that the presence of a given level of
background will decrease the sensitivity of the experiment. As a
first measure, the experiment should be located underground. In
addition, studies are ongoing to determine precisely the loss in
sensitivity for a given level of background, in order to assess to
which extent are low background techniques necessary, including
shielding, vetoing, material radiopurity and gas purification, as
well as offline rejection techniques. Ultimately, an experimental
test in underground location with a prototype is envisioned to
prove that the necessary low background level can actually be
obtained. \item {\bf Radial resolution:} The determination of the
radial coordinate of the interaction point through the measurement
of the time dispersion of the detected charge pulse is a key issue
of the NOSTOS concept. Preliminary results concerning this point
have been already obtained with the prototype detector which is
not equipped with Micromegas and will be presented in the next
section. Work is in progress to assess the final radial resolution
achievable. \item {\bf Scaling up:} Finally, even if good
experimental parameters are demonstrated for small prototypes,
work has to be done to assure that they will be maintained after
the scaling up to the full size required by the NOSTOS experiment.
To this end, an intermediate scale (4 m diameter) prototype is
being designed and will be used in a second stage of the ongoing
development phase.

\item {\bf Electrostatics:} Right now, a metallic rod supports the
central sphere. This provokes a distortion of the electric field,
so that only a third of the volume (opposite to the stick) is
reasonable close to the desired spherical field of (\ref{Efield}).
Two main ideas are being considered to solve this problem:
\begin{itemize}
\item {\bf a)} Use of field shaping rings along the metallic rod
at corresponding potentials. The extremity of the rod, next to the
small sphere, would be made of a resistive cone. \item {\bf b)} A
charging system like the one used in electrostatic accelerators,
using a series of small metallic balls on an insulator chain. Due
to the absence of beam discharging the "terminal", a very small
chain would be sufficient.
\end{itemize}

\end{itemize}

Using a 20 kg tritium source (=200 MCi) the total number of
emitted neutrinos is $6\times 10^{18}\nu/s$. With the TPC filled
with Xe at 1 bar the number of detected neutrinos is about
1000/year, assuming an energy threshold of 100 eV. The use of a
less intense source or cheaper gases like Ar or Ne is possible at
the expense of operation at higher pressures. High pressure TPC
with Micromegas readout have been successfully tested
\cite{Gorodetzky} in the past. They are included in the NOSTOS
development program in order to study the viability of such mode
of operation.


A detailed study based on Monte Carlo simulations with realistic
experimental parameters and with the rigorous treatment of the
neutrino interaction including atomic effects developed in
\cite{Gounaris:2004ji} is currently under way in order to
determine more precisely the sensitivity prospects of this
proposal.

\section{Preliminary results from the first spherical prototype
with a new proportional counter}

A prototype of spherical TPC has been built as a first step
towards the NOSTOS detector, and is currently being used at Saclay
to perform demonstration tests. The spherical vessel is 1.3 m of
diameter and is made of 6 mm thick copper, allowing to hold up to
5 bar of pressure. The first tests were oriented to the assessment
of the tightness of the vessel, so the gas could keep the
sufficient level of purity for right operation. The volume was
pumped by a primary pump followed by a turbo molecular pump,
reaching a level of vacuum below $10^{-6}$ mbar. The outgassing
rate measured was below $10^{-9}$ mbar/s, which allows us to avoid
permanent gas circulation through special cleaning filters and to
operate instead in seal mode.

As mentioned before, a Micromegas-type readout is proposed as
amplification structure in the center of the TPC. Work is in
progress to actually design and build a spherical Micromegas
detector with new technologies. A more conventional alternative
would be to approximate the spherical geometry by a composition of
several flat Micromegas elements. While working independently in
the design of the amplification structure, the first tests with
the spherical vessel were performed using a small spherical
electrode (10 mm diameter) placed in the center of the TPC,
working as a proportional counter.
The signals from this proportional counter are very slow, for only
the ion movement can be observed, and their path is long. The use
of Micromegas, where the electron movement in the 100 $\mu$m gap
will be observed will lead to much shorter pulses. The system
shows a very small capacity, $\sim$1pF, and therefore a very low
level of electronic noise is expected ($< 1000{\rm e}^-$) allowing
for a potentially very low threshold.

\begin{figure}[t]
  \includegraphics[height=.5\textheight]{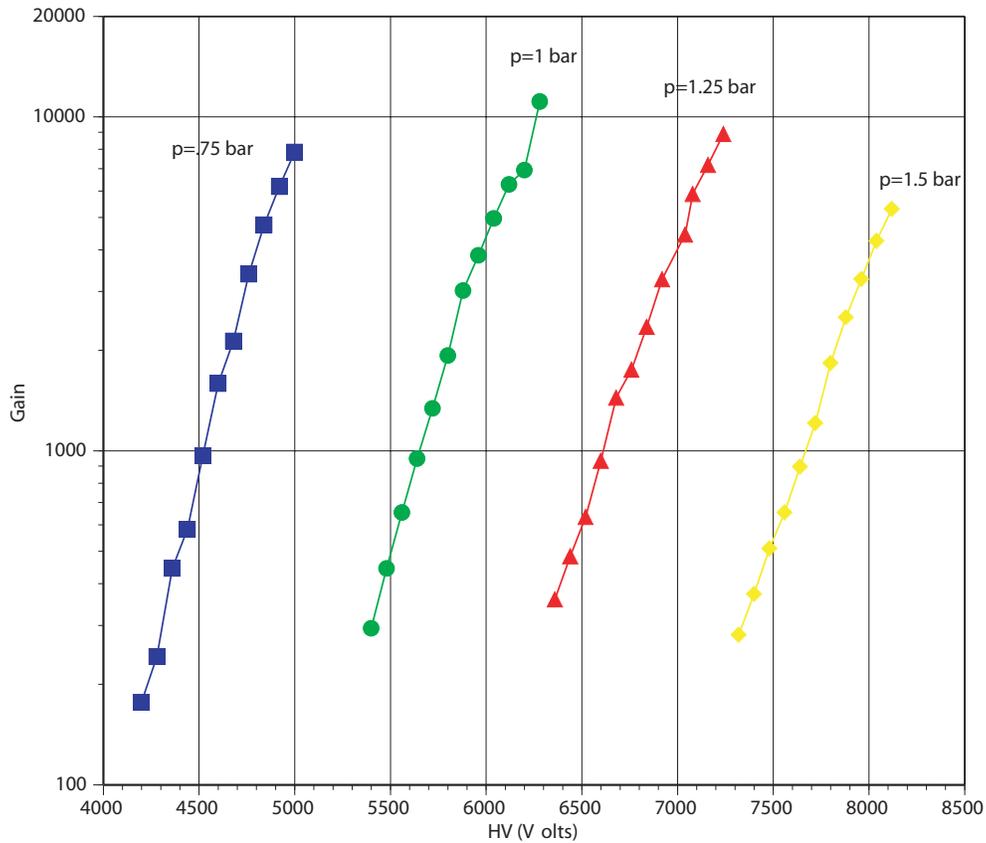}
 \caption{Measured gain as a function of the voltage
 applied, for Ar + 2\% Isobutane at four different pressures.
 \label{gain}}
\end{figure}

\begin{figure}[t]
  \includegraphics[height=.35\textheight]{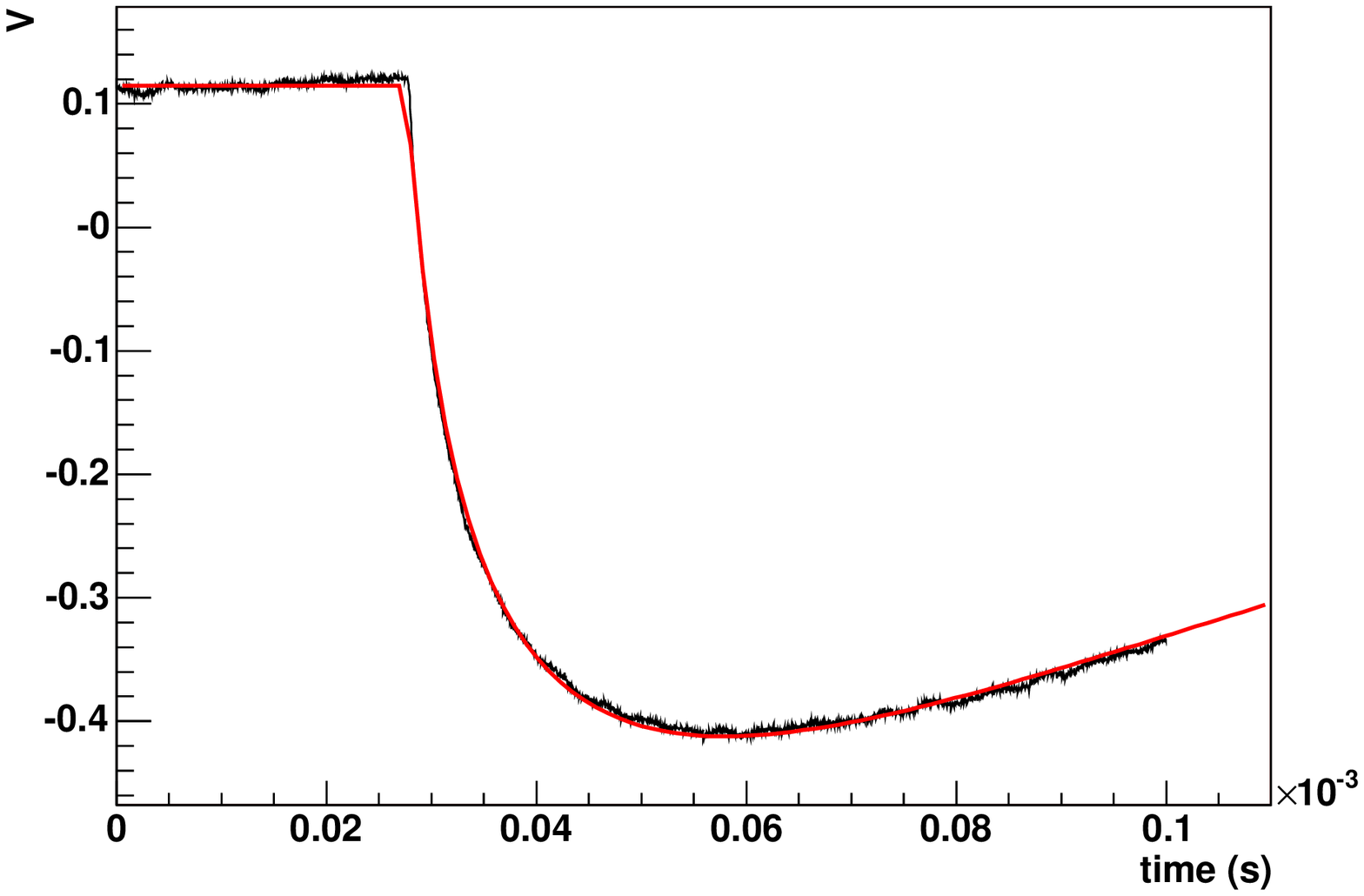}
 \caption{Example of a Fe$^{55}$ pulse.
 \label{pulse}}
\end{figure}

\begin{figure}[t]
  \includegraphics[height=.35\textheight]{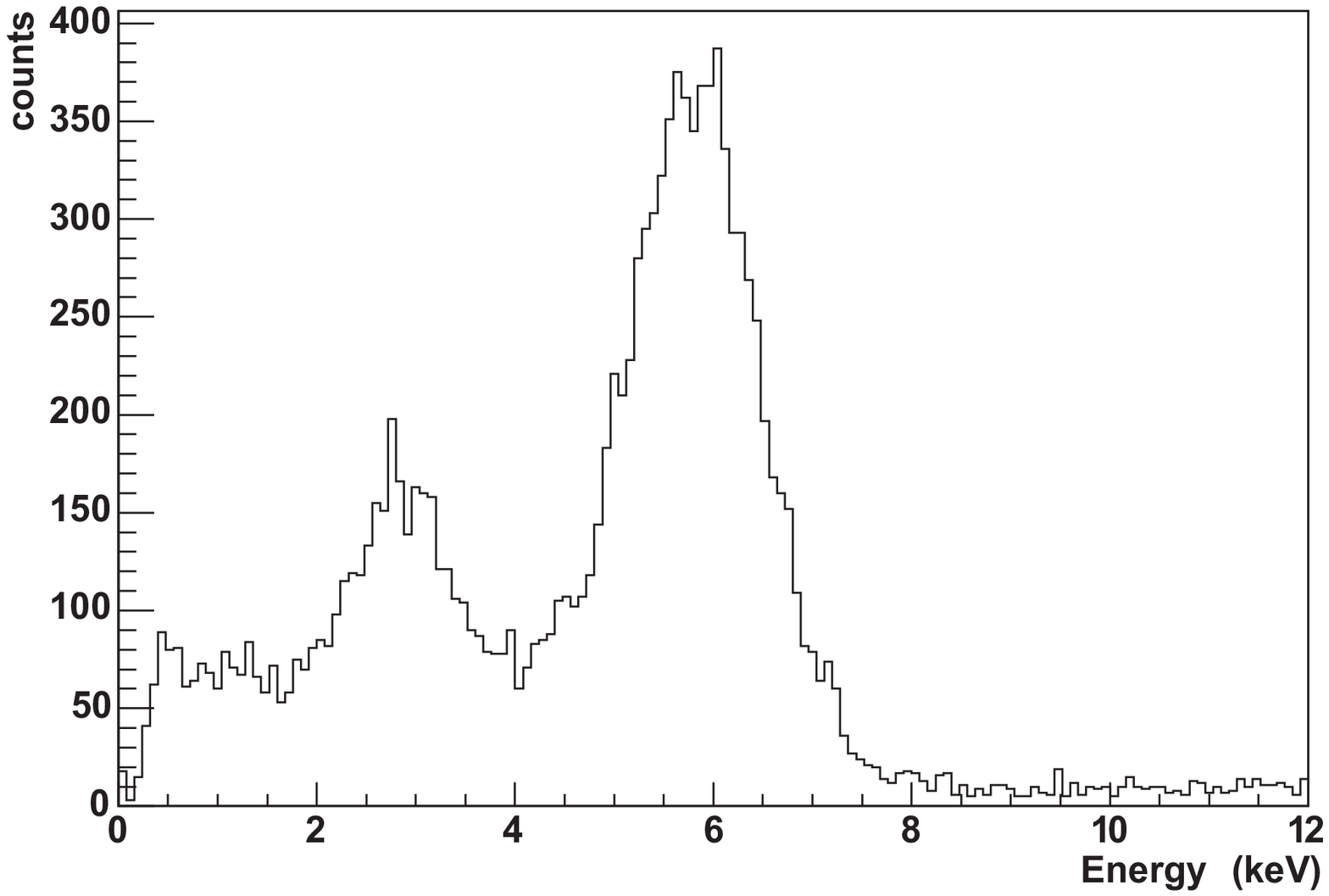}
  \caption{$^{55}$Fe spectrum.
\label{spectrum}}
\end{figure}

%
%
%


So far the prototype has been operated with two different gas
mixtures, namely Ar + 10\% CO$_2$ as well as Ar + 2\% Isobutane;
and at different pressures up to 1.5 bar. These first tests showed
that even with such a simple amplification element, high gains
(above $10^4$) are achieved. Figure \ref{gain} shows the obtained
gain versus voltage curves. Stable operation was tested up to 40
days, without gas circulation (seal mode). This results are very
encouraging but we are not yet in the tritium configuration. Runs
using calibration sources of $^{109}$Cd and $^{55}$Fe or cosmic
rays have been performed.
An example of a $^{55}$Fe 5.9 keV event is shown in Fig.
\ref{pulse}. The remarkable low noise that the baseline of the
pulse of fig. \ref{pulse} shows that thresholds as low as
$\sim$100 eV are already at hand.

To study the drift, the $^{55}$Fe source is introduced inside the
sphere by means of a movable insulant stick which allows us to
place the source at any distance from the inner electrode. Data
taken at different source distances show no evidence of loss of
signal intensity due to electron attachment.
In Figure \ref{spectrum} the recorded energy spectrum of one of
these runs is shown.

\begin{figure}[t]
  \includegraphics[height=.4\textheight]{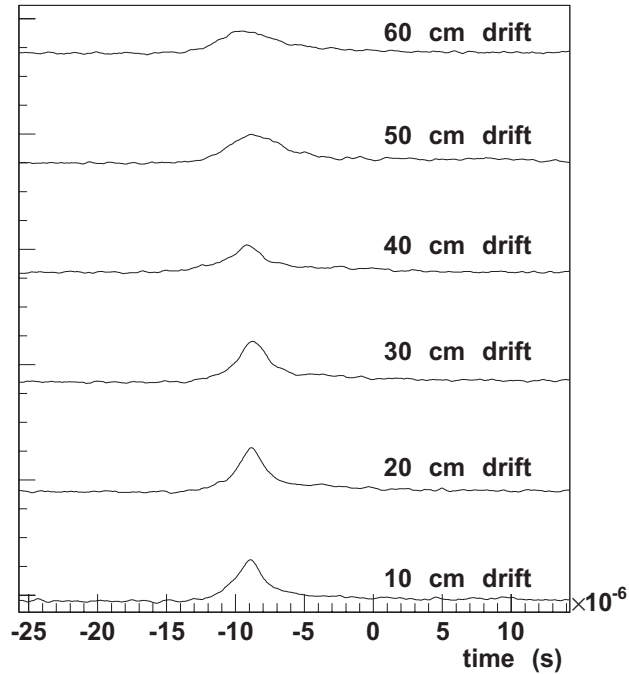}
 \caption{Effect of the diffusion on the deconvolved pulses. Each line
 corresponds to pulses taken with the source at different distances
 from the inner electrode, from 10 cm up to 60 cm in steps of 10 cm.
 \label{diffusion}}
\end{figure}

These data allow the study of the drift properties of the chamber.
Preliminarily, no appreciable electron attachment has been
observed. The main concern at this point of the development phase
is to demonstrate whether the time diffusion of the event can be
measured and the drift distance extracted from it. To this end, as
long as we do not equip the anode with Micromegas, a detailed
Pulse Shape Analysis is being developed based on deconvolution
techniques. Although the work is still under progress, the first
results are encouraging, showing that a 10 cm resolution is
already achievable as shown in Fig.~\ref{diffusion}. The technique
is supposed to unfold the effect of the electronics and the charge
induction from the raw pulse, to arrive to the deconvoluted pulse,
a bare reflection of the temporal pattern of the electron cloud
arriving at the central electrode. In Fig.~\ref{diffusion} various
``template pulses'', each one obtained by averaging 20
deconvoluted pulses, are shown for different positions of the
calibration source. The effect of the diffusion (as the width of
the pulse) is clearly visible.

Due to the mean free path of the $^{55}$Fe X-rays in the gas,
there is an uncertainty in the position of the interactions in
those tests. To overcome this difficulty a new way of calibrating
this effect is being designed. It consists of the use of a
$^{241}$Am source, emitting simultaneously an $\alpha$ and a 60
keV $\gamma$. The $\alpha$ is detected by a small silicon diode
providing the trigger to detect the $\gamma$ interacting in the
sphere. The time delay between the $\alpha$ and the arrival of the
electrons at the small sphere provides an accurate measure of the
drift time.

\section{Additional physics program}

We briefly mention some additional application of the NOSTOS
detector in the domain of neutrino physics.

\subsection{Sensitivity to the neutrino magnetic moment}

Because of the low energy of the incoming neutrinos and the low
energy electron recoils detected in this experiment the
sensitivity for the neutrino magnetic moment is high. The cross
section of the magnetic moment can be written
as\cite{McLaughlin:2003yg}:
\begin{equation}\label{n_coherent}
\left(\frac{d\sigma}{dT}\right)_{EM} =  \sigma_0
\left(\frac{\mu_l}{10^{-12}\mu_B}\right)^2
\frac{1}{T}\left(1-\frac{T}{E_{\nu}}\right)
\end{equation}
Because of the dependance $1/T$ ($T$ is the electron recoil
energy) the sensitivity for the magnetic moment is obviously
higher at low energy. Recent measurements from the MUNU experiment
\cite{Daraktchieva:2003dr} show a limit of 10 $^{-10}$ $\mu_B$ for
the neutrino magnetic moment. Our experiment opens the way to
improve this value by two orders of magnitude.

\subsection{Measurement of the Weinberg angle at low energies}

Another interesting quantity is the Weinberg angle appearing in
standard neutrino cross-sections, which is a function of the
momentum transfer and it has not been measured at such low
transfers. To this end atomic physics experiments, which utilize
the neutral current, have thus far been considered.

By plotting the differential neutrino-electron cross section as a
function of the electron energy we obtain a straight line. We hope
to construct the straight line quite accurately. Thus we can
extract a value of the Weinberg angle both from the slope and the
intercept achieving high precision.

\subsection{Supernova sensitivity. Coherent neutrino scattering}

It is generally believed that the core-collapse supernova
explosion produces a large number of neutrinos and 99\% of  the
gravitational energy is transformed to neutrinos of all types. The
supernova (SN) neutrino flux consists of two main components: a
very short ($<$ 10 msec) pulse of $\nu_e$ produced in the process
of neutronization of the SN matter through the reaction e + p
$\rightarrow $ e + n, which is followed by a longer ($<$ 10 sec)
pulse of thermally produced $\nu_e$, $\nu_\mu$, $\nu_\tau$, and
their antiparticles. Only a small fraction, about 1\%, of the
neutrinos are prompt, while the rest are neutrino-antineutrino
pairs from later cooling reactions. It is expected that spectra of
thermally produced neutrinos are characterized by the different
mean energies: $\nu_e=11$ MeV, $\bar{\nu}_e=16$ MeV,
$\nu_{e,\mu}=25$ MeV. Our idea is to use the large cross section
offered by the coherent neutrino-nucleus cross section for
detecting neutrinos from Super Nova explosions. Coherent
scattering occurs when neutrinos interact with more than one
particle and the amplitudes from the various constituants of the
target add up. The increase of the cross section is proportional
to the square of the number of particles in the target leading to
increased counting rates:
\begin{equation}\label{crosssection}
 \sigma = \frac{G^2N^2E^2}{4\pi}
\end{equation}
where $G$ is the weak coupling constant, $N$ is the number of
neutrons in the target nucleus and $E$ is the neutrino energy. In
order to get advantage of the coherent scattering amplification of
heavy nuclei gases are needed.

For instance, using Xenon as detector target the coherent  cross
sections at  E=25  MeV, the energy that is relevant for Supernova
detection, is  quite  large ($\sigma =1.5\times 10^{-38}$ cm$^2$).
Even at lower energy  (11  MeV) where  the  coherent  cross
sections decreases quadratically  with  energy,  the cross section
is still high. The recoil energy energy is quite low and it takes
a maximum value of 1.5 keV for 11 MeV and 9 keV for 25 MeV
neutrinos. This implies that detector thresholds must be set quite
low with one advantage that backgrounds are highly suppressed
given the narrow time window in which the burst takes place. The
collected energy may be even lower by a significant factor
(quenching factor) and therefore sub-keV detector threshold is
required. For a typical galactic SN explosion the detector used
for the tritium experiment (10 m in radius, p=10 bar of Xenon) the
number of detected neutrinos will exceed 100,000. A possibility to
test the efficiency of detecting coherent neutrino scattering will
be the nuclear reactor. The expected number of neutrino
interactions in the gas volume, from a typical reactor neutrino
flux and spectrum ($10^{13}$ cm$^{-2}$ s$^{-1}$) using a detector
filled with Xenon is about 350/day/Kg\cite{Collar:2000jq}. The
drawback is the very low energy threshold needed since the maximal
recoil energy is 185 eV. Therefore single electron counting is
required imposing a high gain operation of the detector (and a
measurement of the quenching factor of the ionization produced by
the low energy recoils). A 4 m diameter prototype, as foreseen as
a later stage of the development phase, would be a perfect tool
for this measurement, as it could contain 2000 kg of Xe at 10 bar.

\section{Conclusions}

The NOSTOS proposal aims at the detection of low energy ($\sim$few
keV) neutrinos from a strong tritium source, in order to measure
the smaller mixing angle $\theta_{13}$. The new concept of the
spherical TPC is proposed as principle of detection. The very
first results of small prototype in operation in Saclay have been
shown.

\end{document}